# CattleSense - A Multisensory Approach to Optimize Cattle Well-Being

*Srijesh Pillai*
*Department of Computer Science & Engineering*
*Manipal Academy of Higher Education*
Dubai, UAE
srijesh.nellaiappan@dxb.manipal.edu

*M. I. Jawid Nazir*
*Department of Computer Science & Engineering*
*Manipal Academy of Higher Education*
Dubai, UAE
jawid_nazir@manipaldubai.com

*Abstract* — **CattleSense is an innovative application of Internet of Things (IoT) technology for the comprehensive monitoring and management of cattle well-being. This research paper outlines the design and implementation of a sophisticated system using a Raspberry Pi Module 4B, RFID Card Reader, Electret Arduino Microphone Module, DHT11 Sensor, Arduino UNO, Neo-6M GPS Sensor, and Heartbeat Sensor. The system aims to provide real-time surveillance of the environment in which Cows are present and individual Cow parameters such as location, milking frequency, and heartbeat fluctuations. The primary objective is to simplify managing the Cattle in the shed, ensuring that the Cattle are healthy and safe.**

## I. Introduction

The livestock industry plays a crucial role in global agriculture, and the health of cattle directly impacts the productivity of this sector[1]. CattleSense leverages IoT technology to enhance cattle monitoring and management. The system combines various sensors and devices to collect and analyze data, enabling access to data that can be utilized to increase the efficiency of milk quality, seamless tracking of the location of individual cattle, and much more. It also includes the identification of unhealthy Cows and facilitates preventive measures that are triggered by the logic running behind the entire IoT Architecture[2].

## II. Literature Review

In October 2023, we visited a dairy farm as part of an Industrial Visit, wherein we spoke to the Dairy Technologist who was managing the well-being of the cows present on the farm. Based on my discussion with him, we concluded that multiple factors determine the well-being of cows. They are as shown in the below diagram.

Each of these factors in turn affects the behavior of the cow, the quality of milk produced by the cow, and the overall health of the cow.

The Fig. 1 below shows a holistic view of the environmental factors that must be taken into account while considering the well-being of Cows, specifically.

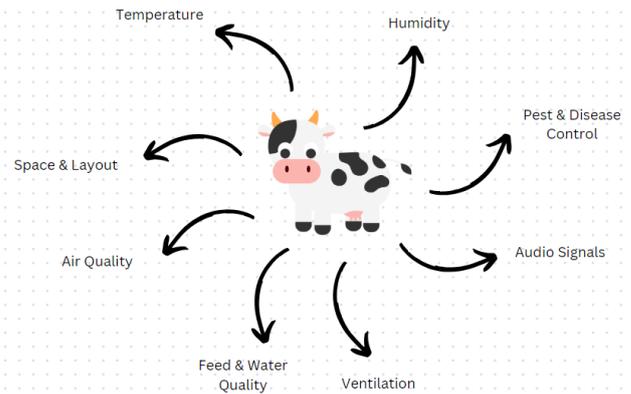

Fig 1. Factors that affect the well-being of the Cow

Previous studies have explored the use of IoT in Agriculture, including the monitoring of livestock[3]. However, there is a gap in the integration of multiple sensors for comprehensive Cattle health monitoring. CattleSense bridges this gap by combining RFID technology for individual cattle identification, environmental sensors such as DHT11 Sensor for monitoring ambient conditions, GPS Sensor for real-time location tracking, and physiological sensors such as Heartbeat Sensor for health assessment and monitoring.

## III. Methodology

The IoT hardware Prototype Design was formed by the identified requirements from the literature review and interviews with the Dairy Technologist of a UAE-based Dairy Product Company. As part of the prototype design process, a comprehensive approach was taken to select appropriate hardware components. The selection criteria included factors such as accuracy, reliability, and compatibility with the intended monitoring objectives.

The Literature Review played a crucial role in identifying relevant IoT hardware elements. Components such as DHT11 Sensors for ambient temperature and humidity measurement, Electret Microphone Modules for audio signal recording, GPS Sensors for live location detection, and RFID Cards for activity monitoring were chosen based on their demonstrated effectiveness in similar applications[2][4].

Considerations were also made regarding the integration of these hardware components to enable seamless data acquisition and processing. Technical feasibility and resource constraints were taken into account to ensure the prototype design aligned with the project objectives and available resources.

Additionally, feedback obtained from the interviews with the Dairy Technologists of the UAE-based Dairy Product Company provided valuable insights into the practical requirements and operational considerations that influenced the selection and design of the IoT hardware prototype. The integration of these insights with the findings from the Literature Review helped refine the Prototype Design to better address the specific needs and challenges of cattle management.

Overall, the Prototype Design phase involved a systematic approach to identify, select, and integrate IoT hardware elements that collectively form the foundation for an innovative and effective solution to address the challenges in cattle management, ensuring a comprehensive approach to livestock well-being[3].

## IV. Design & Implementation

For the implementation of our solution, due to the lack of available resources, we kept our solution simple, using commonly used sensors. This will also assist the Cattle sheds with a lower budget and lack of awareness of such technology to try and adopt solutions of such kind, instead of relying on traditional methods. In our solution as mentioned earlier, we use the DHT11 Sensor for measuring Temperature and Humidity of the Cattle Shed, Heartbeat Sensor to measure the heartbeat of the Cattle, GPS Sensor to track the location of each Cattle in real-time, Electret Arduino Microphone module to capture the Audio Signals, etc.[5].

The Fig. 2 below, shows the Node of the IoT Circuit. It is the one that will be attached to the Cattle at all times. Due to this, it is necessary for it to be of small size, so that it can be fitted easily into the collar of the Cattle, around their neck. The architecture and explanation of the circuit is as follows:

The GPS sensor provides real-time location data. This can help to ensure that the cows are within the premises of the dairy farm and track their real-time location.

The heartbeat sensor tracks the physiological parameters of the cow. This sensor is supposed to touch the cow's skin at all times and monitor the rate of change in the volume of blood, based on which the rate of heartbeat per minute can be computed[6]. A data fusion algorithm processes this information to assess the health status of each cow.

CattleSense integrates the Raspberry Pi Model 4B as the Central Processing Unit, i.e. the Aggregator of the IoT Architecture, coordinating data received from all the IoT components in the circuit, namely, RFID Card Reader, Electret Arduino Microphone Module, DHT11 Sensor, Arduino UNO, Neo-6M GPS Sensor, and Heartbeat Sensor[1]. The collected data is then sent to the cloud for further processing and analytics, which can be viewed by the Administrator of the Cattle Farm[7].

The RFID Card Reader identifies individual cows, while environmental sensors monitor temperature and humidity. For every activity the cow performs, it is important to maintain its frequency, or attendance, to analyze any anomaly in the data. For example, let us say the milking frequency of a certain cow is three times a day. If, through the CattleSense tracking system, the administrator observes that the cow is being milked only two times a day, then the vets can be informed.

Both these circuits can be connected wirelessly using Radio Frequency Transceiver Modules such as NRF24L01, to ensure seamless transmission of data.

The Fig. 3 below, shows Aggregator of the IoT Circuit. It is the one that will collect all the data sensed by the Sensors, and either perform data fusion and data processing at its place, or will send it to the Cloud for advanced analysis. The architecture and explanation of the circuit is as follows:

*Node of the IoT Circuit (installed on every cattle)*

*Aggregator of the IoT Circuit*

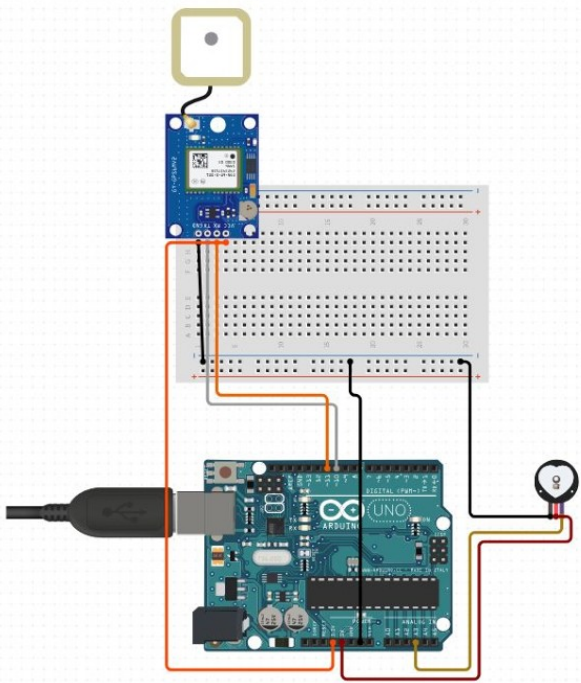

Fig. 2. Circuit diagram of the Node of the IoT Architecture

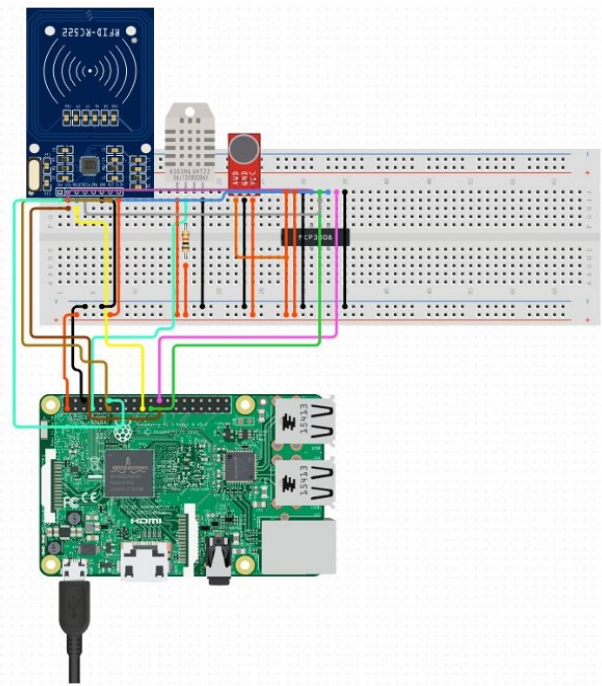

Fig. 3. Circuit diagram of the Aggregator of the IoT Architecture

The communication protocol recommended for communication between the Nodes and the Aggregator of the IoT Circuit is LoRaWAN (Long Range Wide Area Network), due to the reasons as follows:

1) Low Power Consumption: LoRaWAN nodes can operate on low power, extending battery life.
2) Scalability: LoRaWAN networks can scale to accommodate thousands of nodes, making it suitable for farms with a large number of cows.
3) Interoperability: LoRaWAN is an open standard, ensuring interoperability across different vendors' equipment.
4) Data Throughput: While LoRaWAN offers lower data rates compared to some other protocols, it's often sufficient for transmitting periodic sensor data from cows.

One consideration that is to be kept in mind while implementing this Architecture is that since the LoRaWAN communication protocol and the NRF24L01 Transceiver Module operate on different frequency bands, we need to ensure that they are being used for different purposes.

In our proposed solution, the communication between the Nodes and the Aggregator will be using the LoRaWAN communication protocol, as based on the size of the Cattle farm, the distance can be large.

## V. Discussions

To initially understand the actual problem statement of how to maintain a comfortable environment for the cows, we prepared a small questionnaire that we asked the Dairy Technologist at the UAE-based Dairy Product Company. Those questions are as follows:

1) *Temperature Regulation*
    - How do you manage temperature fluctuations to ensure the comfort of the cows, particularly during extreme weather conditions? [6]
    - Are there specific cooling or heating systems in place to maintain optimal temperatures within the cattle shed?

2) *Humidity Control*
    - How do you monitor and manage humidity levels to prevent discomfort and respiratory issues in the cows? [6]
    - What measures are taken to ensure adequate ventilation and airflow to reduce humidity buildup?

3) *Ventilation Systems*
    - Can you describe the ventilation system used in the cattle shed and how it helps maintain air quality and temperature control?
    - How often are ventilation systems inspected and maintained to ensure optimal performance?

4) *Pest and Disease Management*
    - What strategies do you employ to control pests such as flies, ticks, and rodents within the cattle shed?
    - How do you prevent the spread of diseases among the herd, and what biosecurity measures are in place?

5) *Noise Levels and Audio Signals*
    - How do you monitor and control noise levels within the cattle shed to minimize stress and discomfort for the cows?
    - Are there specific measures in place to reduce noise from equipment or other sources that could impact the cow's well-being?

6) *Feed and Water Quality Assurance*
    - How do you ensure the quality and safety of the feed and water provided to the cows?
    - Are there regular tests or inspections conducted to identify and address any potential contaminants or deficiencies?

7) *Air Quality Monitoring*
    - What methods are used to monitor air quality within the cattle shed, particularly regarding ammonia and other harmful gases?
    - How do you mitigate air quality issues to prevent respiratory problems and maintain overall cow health? [8]

8) *Space and Layout Optimization*
    - How is the layout of the cattle shed designed to optimize space and comfort for the cows?
    - Are there considerations for providing sufficient resting areas, feeding stations, and access to water to minimize competition and stress among the herd? [9]

9) *Natural Light and Lighting Systems*
    - How do you incorporate natural light into the design of the cattle shed, and what are the benefits for cow well-being?
    - Are artificial lighting systems used, and if so, how are they optimized to promote cow comfort and productivity?

10) *Environmental Impact Mitigation*
    - What sustainability practices are implemented to minimize the environmental impact of the dairy farm operation, such as waste management and energy efficiency?
    - How do these practices contribute to the overall well-being of the cows and the surrounding ecosystem?

These questions delve into various aspects of environmental conditions and management practices that are crucial for ensuring the well-being and health of cows on a dairy farm. Addressing these factors effectively can help maintain a conducive and healthy environment for the herd, ultimately enhancing their productivity and welfare.

Additionally, we thought of reaching out to a few more Dairy Technologists. For which, we created a small Survey of seven questions. The following graphs depict the responses to the Survey. Fig. 4 below shows a graphical representation of the responses for questions 1 and 2 of the survey. Fig. 5 below shows a graphical representation of the responses for questions 3 and 4 of the survey. Fig. 6 below shows a graphical representation of the responses for questions 5 and 6 of the survey. Finally, Fig. 7 below shows a graphical representation of the responses for question 7 of the survey.

This survey has been filled by Dairy Technologists of different farms situated in multiple geographical locations in India.

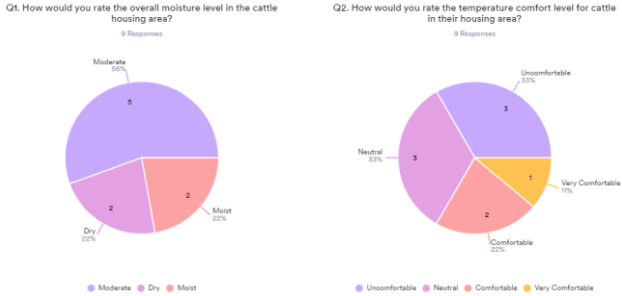

Fig. 4. Graphical representation of all the Survey Responses received from the Dairy Technologists for Questions 1 and 2

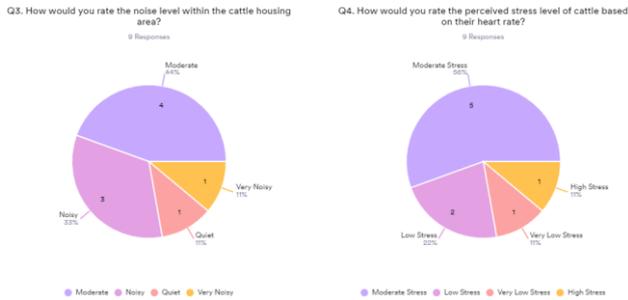

Fig. 5. Graphical representation of all the Survey Responses received from the Dairy Technologists for Questions 3 and 4

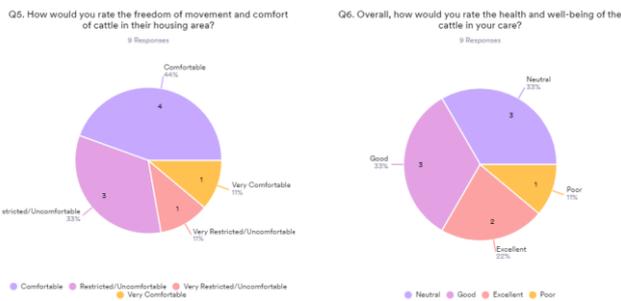

Fig. 6. Graphical representation of all the Survey Responses received from the Dairy Technologists for Questions 5 and 6

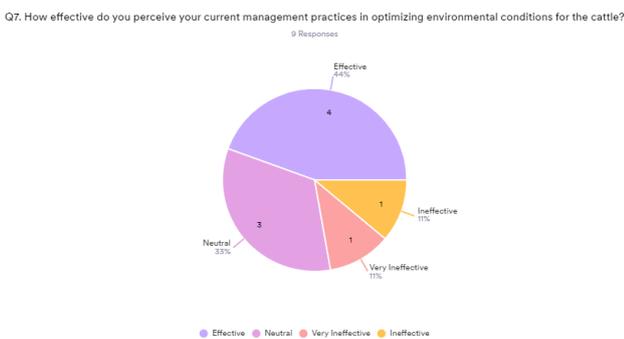

Fig. 7. Graphical representation of all the Survey Responses received from the Dairy Technologists for Question 7

While CattleSense offers a promising solution for cattle monitoring and management, there are several limitations and areas for future research, which are mentioned as follows:

- Scalability: Although LoRaWAN provides scalability for large farms, testing in extremely large-scale operations would be beneficial to evaluate its performance under maximum capacity. Further research could explore advanced network optimization techniques to enhance scalability.

- Adaptability: The system's adaptability to different farming environments needs to be assessed. Each farm may have unique requirements, and future research should focus on customizing the solution to various agricultural settings.

- Data Analytics: The current system focuses on data collection. Enhancing the analytics capabilities to predict health issues and optimize feeding schedules through machine learning algorithms would be valuable.

- Energy Efficiency: While the system is designed for low power consumption, further reducing energy use of the sensors and communication modules would extend battery life and lower operational costs.

- Real-time Processing: Improving the system's real-time data processing capabilities to provide instant alerts and actions in response to critical health or environmental changes.

Addressing these limitations and pursuing the suggested areas for future research will further enhance the robustness and applicability of CattleSense, ultimately leading to more effective and scalable cattle health monitoring solutions.

## VI. RESULTS AND CONCLUSION

Fig. 8 is of the Dashboard that will be seen by the administrator of the Cattle Shed.

On the top-left side is the Humidity of the environment of the Cattle Shed, measured in terms of percentages (%). If the Humidity percentage goes below 30% or above 80%, then external corrective action is to be taken, such as using Humidifiers, Dehumidifiers, Air Conditioners, etc. If the environment is too dry, i.e. the Humidity is very less (below 30%), then it can cause Respiratory Issues, Dehydration, Heat Stress, etc. ultimately leading to decreased milk production. If the environment is too wet, i.e. the Humidity is very high (above 80%), then it can cause fungal growth and infections to the Cattle.

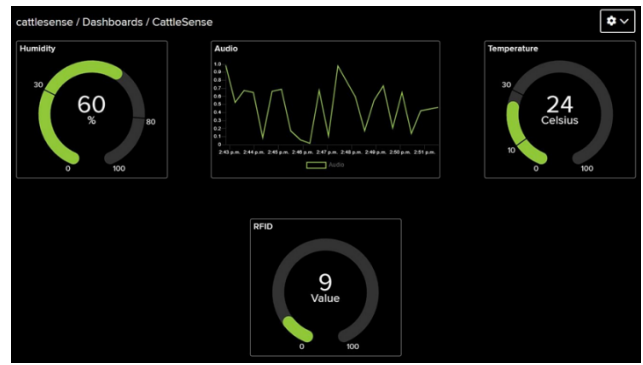

Fig. 8. Dashboard of the CattleSense solution

The top-right side includes the Temperature of the environment measured in terms of Degree Celsius. If the Temperature is not within the range of 10 Degrees Celsius and 30 Degrees Celsius, then external corrective action is to be taken, such as increasing the ventilation of the environment within the shed.

The graphical representation shown in the middle of the dashboard shows the variation of Audio Signals, in terms of Decibels (dB). The recommended range for cows is 35-45 dB, and this condition has been set in the Python code that is running on the Raspberry Pi [10].

The lower-middle area of the Dashboard shows the RFID count for any activity that the Cattle perform on a daily basis within the farm, which can be used for attendance tracking, as mentioned earlier. This count will be reset as and when the activity changes, which is then stored in the Database for further analysis.

Preliminary testing of CattleSense demonstrates its effectiveness in monitoring individual cows and their environments. The system successfully identifies deviations in health parameters, enabling prompt action to prevent the spread of diseases. The real-time data transmission capabilities ensure timely intervention by relevant authorities. Further refinement and field testing are required to validate the system's performance in diverse livestock settings[3].

Below are some ways CattleSense benefits both the Cattle and the Dairy Farm.

A. *Cattle's Perspective[1]*

- Improved Welfare: Enhanced monitoring of temperature and humidity ensures optimal living conditions, promoting the health and well-being of the cattle.
- Timely Intervention: Real-time data acquisition enables swift responses to environmental changes, allowing for prompt interventions in case of unfavourable conditions.
- Enhanced Productivity: The holistic approach to monitoring creates an environment conducive to increased productivity and overall better performance.

The Fig. 9 below is of the output of the GPS Sensor, which is attached to each Cattle. The output in the figure is displayed in terms of Pole Coordinates including the Latitude, Longitude, and Altitude of the Cow. Hence, with information on the boundaries of the Dairy Farm, the administrator can make sure that the Cattle are within the premises of the farm.

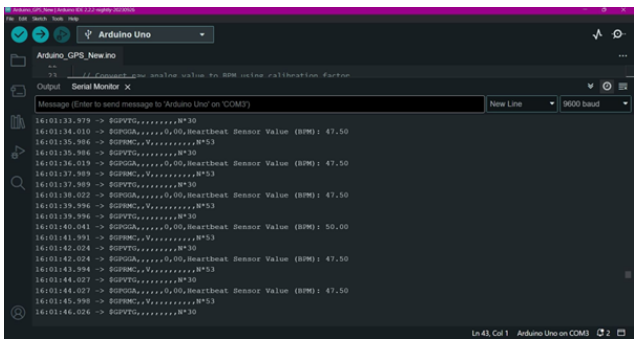

Fig. 9. Output of the GPS Sensor and Heartbeat Sensor in the Arduino IDE

B. *Dairy Farm's Perspective[11]*

- Operational Efficiency: The project facilitates streamlined operations through automated surveillance, reducing the manual effort required for monitoring environmental parameters.
- Data-Driven Decision Making: Access to real-time data on the Interactive Dashboard empowers the dairy company to make informed decisions, optimizing resource allocation and management strategies.
- Healthier Livestock[3]: Proactive monitoring helps prevent potential health issues, reducing veterinary costs and improving the overall health of the cattle, leading to better-quality dairy products.

In conclusion, CattleSense presents a novel approach to cattle health monitoring through the integration of IoT technologies. This system has the potential to revolutionize livestock management, promoting early detection of health issues and preventing the spread of diseases within cattle populations. As agriculture continues to embrace technology, CattleSense stands at the forefront of ensuring the well-being of livestock for a sustainable and efficient farming future.